\documentstyle[aps,12pt,epsfig,amssymb]{revtex}

\begin{document}
\baselineskip=0.7cm
\newcommand{\ini}{\begin{equation}}
\newcommand{\fin}{\end{equation}}
\newcommand{\inir}{\begin{eqnarray}}
\newcommand{\finr}{\end{eqnarray}}
\newcommand{\inif}{\begin{figure}}
\newcommand{\finf}{\end{figure}}
\newcommand{\bc}{\begin{center}}
\newcommand{\ec}{\end{center}}
\def\ol{\overline}
\def\pa{\partial}
\def\ra{\rightarrow}
\def\ts{\times}
\def\df{\dotfill}
\def\bs{\backslash}
\def\dg{\dagger}

$~$

\hfill DSF-31/2000

\vspace{1 cm}

\centerline{\LARGE{Leptogenesis and neutrino parameters}}

\vspace{1 cm}

\centerline{\large{D. Falcone and F. Tramontano}}

\vspace{1 cm}

\centerline{Dipartimento di Scienze Fisiche, Universit\`a di Napoli,}
\centerline{Complesso di Monte Sant'Angelo, Via Cintia, Napoli, Italy}

\centerline{{e-mail: falcone@na.infn.it; tramontano@na.infn.it}}

\vspace{1 cm}

\begin{abstract}

\noindent
We calculate the baryonic asymmetry of the universe in the
baryogenesis-via-leptogenesis framework, assuming first a quark-lepton symmetry
and then a charged-neutral lepton symmetry. We match the results with
the experimentally favoured range. 
In the first case all the oscillation
solutions to the solar neutrino problem, except the large mixing matter
solution, can lead to the allowed range, but with fine tuning of the
parameters.
In the second case the general result is quite similar.
Some related theoretical hints are discussed.
 
\end{abstract}

\newpage

\section{Introduction}

\noindent
Strong indications for nonzero neutrino mass and mixing come from solar and
atmospheric neutrino experiments. In fact, if interpreted in terms of neutrino
oscillations, such experiments, together with the tritium beta decay endpoint,
imply small neutrino masses \cite{bww}.

In the minimal standard model (MSM) the neutrino is massless
because there are no right-handed neutrino singlets and there is no Higgs
scalar triplet. The simplest way to
get a mass for the neutrino field is by adding the right-handed state
$\nu_R$, the analogue of the quark state $u_R$ in the leptonic sector,
in which case it becomes possible to build both a Dirac mass term
$m_{\nu} \ol{\nu}_L \nu_R$ and a Majorana mass term
$(1/2)m_R \ol{\nu}^c_L \nu_R$ for the right-handed neutrino.
The Dirac mass $m_{\nu}$ is expected to be of the same order of magnitude of
the quark or charged lepton masses, while the Majorana mass $m_R$ is not
constrained. A popular mechanism to obtain a very small neutrino mass is
the seesaw mechanism \cite{ss}, where the right-handed neutrino mass is very
large and as a consequence a very light left-handed Majorana neutrino appears,
with a mass $m_L \simeq m_{\nu}^2/m_R$.

The MSM plus the right-handed neutrino (which we would like to call SM)
is also a minimal
scenario to produce a baryonic asymmetry in the universe, according to the
Fukugita-Yanagida baryogenesis-via-leptogenesis mechanism \cite{fy,luty}.
In this framework the out-of-equilibrium decays of right-handed neutrinos
generate a leptonic asymmetry which is partially transformed into a baryonic
asymmetry by electroweak sphaleron processes \cite{krs}.

The baryonic asymmetry depends on both the Dirac and the right-handed
Majorana neutrino mass matrices. Therefore, assuming a quark-lepton symmetry
or a charged-neutral lepton symmetry, we should be
able to determine the value of the baryonic asymmetry, and to match it with the
experimental
bounds coming from nucleosynthesis in the standard big bang theory. This is
the main subject of the present paper, already discussed by several authors
\cite{bpp,bb,gks,no}. However, our approach is quite different,
more general and direct.
We scan over the neutrino parameter space,
using several forms for the Dirac mass matrices,
and taking into account the vacuum
and matter solutions to the solar neutrino problem.
A graphical representation of the results is given, from which one can
eventually infer approximate bounds on neutrino parameters. Both the
nonsupersymmetric (SM) and the supersymmetric (SSM) cases are considered.

Section II is about neutrino oscillation data, from which one may obtain
light neutrino masses and mixings. Section III deals with
the quark-lepton symmetry, which allows to get the Dirac and heavy neutrino mass
matrices. In section IV, after a short collection of the
relevant formulas of the baryogenesis-via-leptogenesis mechanism, the
calculation of the baryonic asymmetry is carried out, based on the content
of sections II and III.
In section V the same calculation is done assuming a charged-neutral lepton
symmetry. Finally, in section VI, we give our conclusions and a
brief discussion.

\section{Neutrino parameters}

\noindent
From the phenomenological point of view the baryonic asymmetry
also depends on which solution for solar neutrinos is taken into account.
Therefore, in this section
we summarize the neutrino oscillation data that we will use in our analysis.
For atmospheric neutrinos the best fit is \cite{kscho}
$$
\Delta m^2_a=3.5 \ts 10^{-3} \text{eV}^2
$$
$$
\sin^2 2 \theta_a=1.0,
$$
that is maximal mixing.
For solar neutrinos we have three matter (MSW) solutions \cite{bks}:
the small mixing angle (SMA)
$$
\Delta m^2_s=5.4 \ts 10^{-6} \text{eV}^2
$$
$$
\sin^2 2 \theta_s=0.006,
$$
the large mixing angle (LMA)
$$
\Delta m^2_s=1.8 \ts 10^{-5} \text{eV}^2
$$
$$
\sin^2 2 \theta_s=0.76,
$$
and the low-$\Delta m^2$ (LOW) solution
$$
\Delta m^2_s=7.9 \ts 10^{-8} \text{eV}^2
$$
$$
\sin^2 2 \theta_s=0.96.
$$
Moreover, we also have the vacuum oscillation (VO) solution \cite{bks}
$$
\Delta m^2_s=8.0 \ts 10^{-11} \text{eV}^2
$$
$$
\sin^2 2 \theta_s=0.75.
$$
The latest day-night and spectral data favour the LMA and LOW solutions,
but do not exclude the others \cite{ellis}.
A further information on neutrino oscillations comes from the CHOOZ experiment
\cite{chooz} which gives
$\sin \theta_c \lesssim 0.16$ for $\Delta m^2_c > 1 \ts 10^{-3} \text{eV}^2$.

Therefore, neutrinos do have masses and mixings, and a unitary matrix
$U_{\alpha i}$
($\alpha=e,\mu,\tau; i=1,2,3$) relates the mass eigenstates $\nu_i$ to the weak
eigenstates $\nu_{\alpha}$,
$$
\nu_{\alpha L}=\sum_i U_{\alpha i} \nu_{i L}.
$$
It is clear that $\Delta m^2_s \ll \Delta m^2_a$. According to
ref. \cite{pet} we assume
$$
\Delta m^2_s= m_2^2-m_1^2,~\Delta m^2_a= m_3^2-m_1^2
$$
where the numbering corresponds to the family index. Moreover, we work with the 
hierarchical spectrum of light neutrinos, $m_1 \ll m_2 \ll m_3$. Then,
$m_3^2 \simeq \Delta m^2_a$, $m_2^2 \simeq \Delta m^2_s$, and for $m_1$ we take
$10^{-4}m_2 < m_1 < 10^{-1} m_2$.

The mixing matrix $U$ (the MNS matrix \cite{mns}) can be written as the
standard parametrization of the CKM matrix (including one phase $\delta'$)
times a diagonal phase matrix $D=\text{diag}(e^{i \varphi_1},e^{i \varphi_2},1)$
\cite{bgg,gg}.
Hence, it depends on three angles and three phases.
From neutrino oscillation data we can determine the three angles \cite{pet,bgg}.
For $|U_{e3}|$, related to the result of the CHOOZ experiment, we use the bound
$$
|U_{e3}| \le 0.2,
$$
while $U_{e2}$ and $U_{\mu 3}$ are obtained from the best fits of
atmospheric and solar neutrinos.
Then we are left with five
free neutrino parameters: $|U_{e3}|$, $\delta=\arg (U_{e3})$, $m_1$,
$\varphi_1$, $\varphi_2$.
Choosing values for the free parameters leads to a complete
determination of light masses and the mixing matrix $U$.
These will be used in the following section,
together with the quark-lepton symmetry, to obtain the heavy neutrino
mass matrix.

\section{Seesaw mechanism with quark-lepton symmetry}

\noindent
The Lagrangian for the relevant lepton sector is
(for simplicity we do not write the 1/2 factor in the Majorana terms)
\ini
{\cal L}=\ol{e}_L M_e e_R+\ol{\nu}_L M_{\nu} \nu_R+g \ol{\nu}_L e_L W+
\ol{\nu}^c_L M_R \nu_R
\fin
where $M_e$ is the mass matrix of charged leptons, $M_{\nu}$ is the mass matrix
of Dirac neutrinos, and $M_R$ the mass matrix of right-handed Majorana
neutrinos. The effective Lagrangian of the seesaw mechanism is
\ini
{\cal L}_{ss}=\ol{e}_L M_e e_R+\ol{\nu}_L M_L \nu^c_R+g \ol{\nu}_L e_L W+
\ol{\nu}^c_L M_R \nu_R
\fin
with the light neutrino mass matrix $M_L$ given by
\ini
M_L=-M_{\nu} M_R^{-1} M_{\nu}^T.
\fin
Setting
\ini
U_L M_L U_L^T=D_L,~U_{eL} M_e U_{eR}^{\dg}=D_e,
\fin
where $D_L$, $D_e$ are diagonal matrices, we obtain the MNS matrix as
\ini
U=U_L U_{eL}^{\dg}.
\fin
Inverting eqn.(3) we get the heavy neutrino mass matrix
\ini
M_R=-M_{\nu}^T M_L^{-1} M_{\nu}.
\fin
Now, assuming a quark-lepton symmetry, we take the pair of hermitian matrices
\ini
M_{\nu}=\frac{m_{\tau}}{m_b}
\left( \begin{array}{ccc}
        0 & 0 & \sqrt{m_u m_t} \\
        0 & m_c & 0 \\
        \sqrt{m_u m_t} & 0 & m_t
\end{array} \right),
\fin
\ini
M_e=\frac{m_{\tau}}{m_b}
\left( \begin{array}{ccc}
        0 & \sqrt{m_d m_s}\text{e}^{\text{i} \alpha} & 0 \\
        \sqrt{m_d m_s}\text{e}^{-\text{i} \alpha} & -3 m_s & \sqrt{m_d m_b} \\
        0 & \sqrt{m_d m_b} & m_b
\end{array} \right),
\fin
with one phase $\alpha=\pi/2$ in $M_{e12}$.
These lepton mass matrices are obtained in the following way.
We take the five texture zero model for the quark mass matrices $M_u$
and $M_d$ from ref.\cite{cf},
which studies the phenomenologically viable textures. The quark
mass matrices are related to the lepton mass matrices by an approximate running
factor $m_{b}/{m_{\tau}}$ from the high scale where the quark-lepton symmetry
should hold \cite{acpr},
and in addiction a factor $-3$ is included in $M_{e22}$
in order to have a good relation between charged lepton and down quark masses
\cite{gj}.
For five texture zeros the matrices $M_u$, $M_d$ lead to the simple meaningful
relations
\ini
V_{us} \simeq \sqrt{\frac{m_d}{m_s}},~
V_{cb} \simeq \sqrt{\frac{m_d}{m_b}},~
V_{ub} \simeq \sqrt{\frac{m_u}{m_t}}.
\fin
In this way, by means of eqns.(4)-(8),
we can calculate $M_R$ and then its eigenvalues $M_1,M_2,M_3$ \cite{df}.
The quark-lepton symmetry is usually obtained within unified theories such as
$SU(5)$ and mostly $SO(10)$, where quarks and leptons belong to the same
multiplets. In particular, the factor $-3$ in $M_{e22}$ is due to suitable
Yukawa couplings of these multiplets with the {\bf 45} (in $SU(5)$) or {\bf 126}
(in $SO(10)$) Higgs representations.
However, here we can also regard the quark-lepton symmetry as a
phenomenological feature.

\section{The baryonic asymmetry}

\noindent
A baryonic asymmetry can be generated from a leptonic asymmetry \cite{fy}.
In order to study this baryogenesis-via-leptogenesis mechanism
we diagonalize $M_e$:
$$
{\cal L}'=\ol{e}_L D_e e_R+\ol{\nu}_L M_{\nu}' \nu_R+g \ol{\nu}_L e_L W+
\ol{\nu}^c_L M_R \nu_R,
$$
where
\ini
M_{\nu}'=U_{eL} M_{\nu},
\fin
and also $M_R$ by means of $U_R M_R U_R^T= D_R$:
$$
{\cal L}''=\ol{e}_L D_e e_R+\ol{\nu}_L M_{\nu}'' \nu_R+g \ol{\nu}_L e_L W+
\ol{\nu}^c_L D_R \nu_R
$$
where
\ini
M_{\nu}''=M_{\nu}' U_R^T \equiv M_D.
\fin
Due to electroweak sphaleron effect,
the baryonic asymmetry $Y_B$ is related to the leptonic asymmetry $Y_L$ by
\cite{ht}
\ini
Y_B=a Y_{B-L}=\frac{a}{a-1} Y_L
\fin
with
$$
a=\frac{8N_f+4N_H}{22N_f+13 N_H},
$$
where $N_f$ is the number of families (three) and $N_H$ the number of Higgs
doublets (one in the SM and two in the SSM; $a \simeq 1/3$ in both cases).
Remember that
$$
Y_B =\frac{n_B-n_{\ol{B}}}{7.04 n_{\gamma}}
$$
where $n_{B,\ol{B},\gamma}$ are number densities.
The leptonic asymmetry can be written as \cite{luty}
\ini
Y_L=d ~\frac{\epsilon_1}{g^*}
\fin
where, in the SM, the CP-violating asymmetry $\epsilon_1$ is
given by \cite{crv,bp}
\ini
\epsilon_1=\frac{1}{8 \pi v^2 (M_D^{\dg} M_D)_{11}}\sum_{j=2,3}
$Im$ [(M_D^{\dg} M_D)_{j1}]^2 f \left( \frac{M_j^2}{M_1^2} \right),
\fin
with
$$
f(x)=\sqrt{x} \left[1-(1+x) \ln \frac{1+x}{x}-\frac{1}{x-1} \right],
$$
$g^*(\text{SM})=106.75$; $v$ is the VEV of the SM Higgs doublet.
In the SSM, $v \ra v \sin \beta$, $f(x) \ra g(x)$,
$g^*(\text{SSM})=228.75$,
$$
g(x)=-\sqrt{x} \left[ \ln \frac{1+x}{x} +\frac{2}{x-1} \right],
$$
and a factor $4$ is included in $\epsilon_1$ \cite{crv}, due to more decay
channels.
For a hierarchical spectrum of heavy neutrinos
$f \simeq -3M_1/2M_j$, $g \simeq -3M_1/M_j \simeq 2f$,
with a very good accuracy.
Eqn.(14) arises from the interference between the tree level and one loop
decay amplitudes of the lightest heavy neutrino, and includes vertex and
self-energy corrections. The latter may be dominant if $M_1$ and $M_j$
are nearly equal, so that an enhancement of the asymmetry may occur.

A good approximation for $d$, the dilution factor,
is inferred from refs. \cite{kt,ap,fp}:
\ini
d=(0.1~k)^{1/2} \exp[-(4/3)(0.1~k)^{1/4}]
\fin
for $k \gtrsim 10^6$,
\ini
d=0.24/k(\ln k)^{3/5}
\fin
for $10 \lesssim k \lesssim 10^6$, and
\ini
d=1/2 k,~~d=1
\fin
for $1 \lesssim k \lesssim 10$, $0 \lesssim k \lesssim 1$, respectively,
where the parameter $k$ is
\ini
k = \frac{M_P}{1.7 v^2 32 \pi \sqrt{g^*}}\frac{(M_D^{\dg} M_D)_{11}}{M_1},
\fin
and $M_P$ is the Planck mass. In the SSM the critical value $10^6$ for $k$ is 
lowered, but in our calculation $k$ remains always much smaller. 
The presence of the dilution factor in eqn.(13) takes into account
the washout effect
produced by inverse decay and lepton number violating scattering.

We make a random extraction of the free neutrino parameters
for a total of 8000 points and we plot $Y_B$ versus such parameters.
As expected, about 4000 points give a negative $Y_B$.
Only $|U_{e3}|$ and $\delta$ show a major effect and the results
for the five texture zero model
are presented in figs. 1-4, according
to the four different solar neutrino solutions. Changing the other parameters,
in particular $U_{e2}$ and $m_2$, within the allowed experimental limits,
does not affect the general result. 
Since the favoured range for the baryonic asymmetry is \cite{osw}
$$
Y_B=(1.7 \div 8.9) \ts 10^{-11},
$$
one can look at the region of $Y_B$ between $10^{-11}$ and $10^{-10}$.
The SMA, VO and LOW solutions can produce the required amount of baryonic
asymmetry, but with fine tuning of the parameters.
Notice that we plot $\text{Log}_{10} Y_B$, which is negative.
However, the trend is clear. For example, we find
the phase $\delta$ tuned around $\pi-\alpha$, which corresponds to
$\delta'=\alpha$, $\sin \theta_{e3}<0$. Moreover, according to ref.\cite{no},
we find an enhancement
of the asymmetry for $|U_{e3}| \simeq V_{D12} U_{\mu 3}$, where
$V_D=U_{\nu L} U_{eL}^{\dg}$ is the mixing matrix in the Dirac sector (the
analogue of $V_{CKM}=V_u V_d^{\dg}$).
Since $V_{D12}\simeq (1/3)\sqrt{m_d/m_s}  \simeq 0.07$,
we get the maximum of $Y_B$ around $|U_{e3}| \simeq 0.07 \cdot 0.7 \simeq 0.05$.
In a similar way one can explain why the SMA solution shows an enhancement,
contrary to the LMA solution. In fact, the further condition is
$U_{e2} \simeq V_{D12} U_{\mu 2}$, which is compatible with the
SMA but not with the LMA.
The two conditions correspond to the decoupling of $M_1$ from $m_3$ and $m_2$,
respectively.
Note that in ref.\cite{no} $V_{D12} \simeq 0.21$
because there the factor $-3$ in our $M_{e22}$ is absent. In this case one has
a different enhancement value for $U_{e2}$, $|U_{e3}|$, so that the SMA is also
excluded. The presence of the factor $-3$ allows the SMA to be reliable
for leptogenesis, for the matrix texture (7),(8).

In the supersymmetric case the calculated
baryonic asymmetry is increased by a factor nearly 6.
In fact, going from the SM to the SSM, there is a factor 4 due to
$\epsilon_1$, a factor $1/\sqrt{2}$ due to ${g^*}$ (for $d \sim 1/k$)
and a factor 2 due to $g(x)$: $4 \cdot (1/\sqrt{2}) \cdot 2 \simeq 6$.
Hence, in the present context, the SSM works better for leptogenesis,
with respect to the SM.

To test the dependence of $Y_B$ on matrix texture, let us consider a second
pair of hermitian matrices \cite{cf}, with four texture zeros, which is
formed by the same matrix $M_e$ as in eqn.(8), but with a further phase
$\pi/2$ in $M_{e23}$, and
\ini
M_{\nu}=\frac{m_{\tau}}{m_b}
\left( \begin{array}{ccc}
        0 & \sqrt{m_u m_c} & 0 \\
        \sqrt{m_u m_c} & m_c & \sqrt{m_u m_t} \\
        0 & \sqrt{m_u m_t} & m_t
\end{array} \right).
\fin
Note that this form for $M_{\nu}$ has entries 1-2 and 2-3 filled in, with
respect to matrix (7), while $M_{\nu 13}=0$. Results useful are in figs. 5, 6,
for the SMA, LOW and VO solutions, while the LMA gives very small asymmetry
as in the foregoing case. For SMA, LOW to work one has $|U_{e3}| \simeq 0.01$,
and for VO also $\delta \simeq \pi/2$.

\section{The case of a charged-neutral lepton symmetry}

\noindent
If there is a charged-neutral lepton symmetry,
the Dirac neutrino mass matrix is related to the charged lepton
mass matrix rather than to the up quark mass matrix:
$M_{\nu} \sim M_e$, with
\ini
M_e=\left( \begin{array}{ccc}
        0 & \sqrt{m_e m_{\mu}} & 0 \\
        \sqrt{m_e m_{\mu}} & m_{\mu} & \sqrt{m_e m_{\tau}} \\
        0 & \sqrt{m_e m_{\tau}} & m_{\tau}
\end{array} \right),
\fin
for example (four texture zeros).
This form for $M_e$ is obtained by analogy to $M_d$ in ref.\cite{cf}. There are
some theoretical (left-right) models \cite{bdm}
with a charged-neutral lepton symmetry,
along with an up-down symmetry. However, again, we can also assume it as a
phenomenological hypothesis.
The position of phases is somewhat arbitrary. We put phases $\pi/2$ in
$M_{e12}$, $M_{e23}$ in order to have a mixing in the Dirac sector similar
to the previous case.
The value of the baryonic asymmetry $Y_B$ is quite analogous to the case of a
quark-lepton symmetry, see figs. 7, 8. However, the maximum level of asymmetry
is now reached for $|U_{e3}| \simeq 0$.

Also for this case, we have checked the dependence on matrix texture by using
the hermitian matrices formed by six texture zeros
$$
M_{\nu} \sim M_e \sim \left( \begin{array}{ccc}
        0 & 0 & \sqrt{m_e m_{\tau}} \\
        0 & m_{\mu} & 0 \\
        \sqrt{m_e m_{\tau}} & 0 & m_{\tau}
\end{array} \right),
$$
with a single phase $\pi/2$ in $M_{e13}$, but we have found no relevant
difference with the asymmetry generated by matrix (20).

\section{Conclusion and discussion}

\noindent
The baryonic asymmetry $Y_B$ has been calculated using a random extraction for
five of the nine neutrino parameters (three light masses; three angles and
three phases in the mixing matrix) and assuming quark-lepton symmetry or
charged-neutral lepton symmetry for the Dirac mass matrices.
Other parameters have been checked.
As a result, for quark-lepton symmetry, we find that the
SMA, VO, and LOW  solutions for solar neutrinos are able to generate
enough asymmetry, especially in the supersymmetric case, but with
fine tuning and selected values of the parameters $|U_{e3}|$ and $\delta$.
For charged-neutral lepton symmetry the general results are
similar.

Let us discuss some related theoretical issues.
Unified theories such as $SO(10)$,
or left-right models such as $SU(3)_c \ts SU(2)_L \ts SU(2)_R \ts U(1)_{B-L}$,
naturally contain heavy Majorana neutrinos, generated at the unification or
left-right scale \cite{mp},
but also contain other particles, for example additional gauge bosons.
Usually these particles are much heavier than the lightest heavy Majorana
neutrino, so that they are sufficiently decoupled from the leptogenesis process,
as confirmed in ref.\cite{cfl}.
In this way, the idea of baryogenesis through leptogenesis may be
attractive also within unified or left-right models.
The VO solution with quark-lepton symmetry gives the scale of $M_R$ around the
Planck mass, while the SMA and LMA solutions
give the scale of $M_R$ near the unification scale ($10^{16}$ GeV)
\cite{df}. Also the LOW solution may be consistent with the unification scale.
Thus, it is hard to reconcile the VO solution with quark-lepton symmetry,
in the context of unified theories, whereas for the SMA and perhaps the
LOW solutions
this is possible. We point out that if the LMA is the right solution to the
solar neutrino problem, then the framework used in this paper
does not work
for leptogenesis. If the VO solution is right, then a good amount of
leptogenesis can be obtained, but with fine tuning and outside normal
unified models. The SMA and LOW solutions may be consistent with both
unified theories and leptogenesis bounds, for selected values of
the complex parameter $U_{e3}$.

In the case of charged-neutral lepton symmetry the VO solution gives $M_R$
around the unification scale, while the SMA and LMA solutions give $M_R$
near the intermediate (left-right) scale ($10^{12}$ GeV) \cite{df}.
The LOW solutions lies between the two.
Hence, the VO solution may be consistent with both the
unification scale and leptogenesis,
and the SMA with both the intermediate scale and leptogenesis.
Since the quark-lepton symmetry is natural in unified models, while the
charged-neutral (up-down) symmetry is natural in left-right models,
in the present context
the preferred solution for solar neutrinos could be the SMA. However,
neutrino data slightly favour the LMA solution \cite{flmp}.
A possible alternative for the baryogenesis-via-leptogenesis mechanism to work
more extensively is by means of horizontal symmetries \cite{ho}.

\section*{Acknowledgements}

We thank G. Covone for a quick help to Mathematica and F. Buccella for comments
on the manuscript.

\newpage

\begin{figure}[ht]
\begin{center}
\epsfig{file=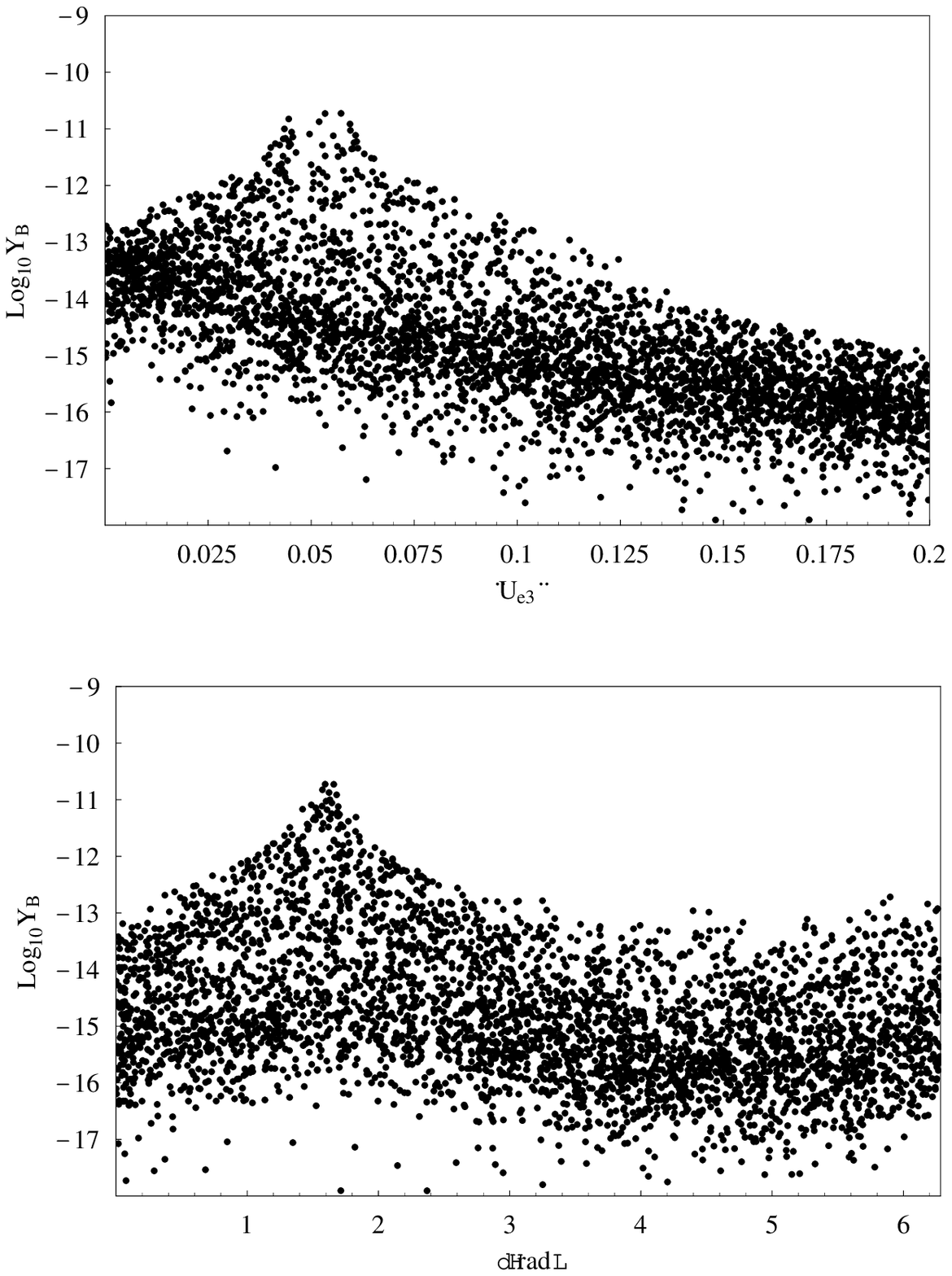,height=16cm}
\caption{The baryonic asymmetry $Y_B$ vs. $|U_{e3}|$ and $\delta$ for SMA,
quark-lepton symmetry, five texture zeros}
\end{center}
\end{figure}

\newpage

\begin{figure}[ht]
\begin{center}
\epsfig{file=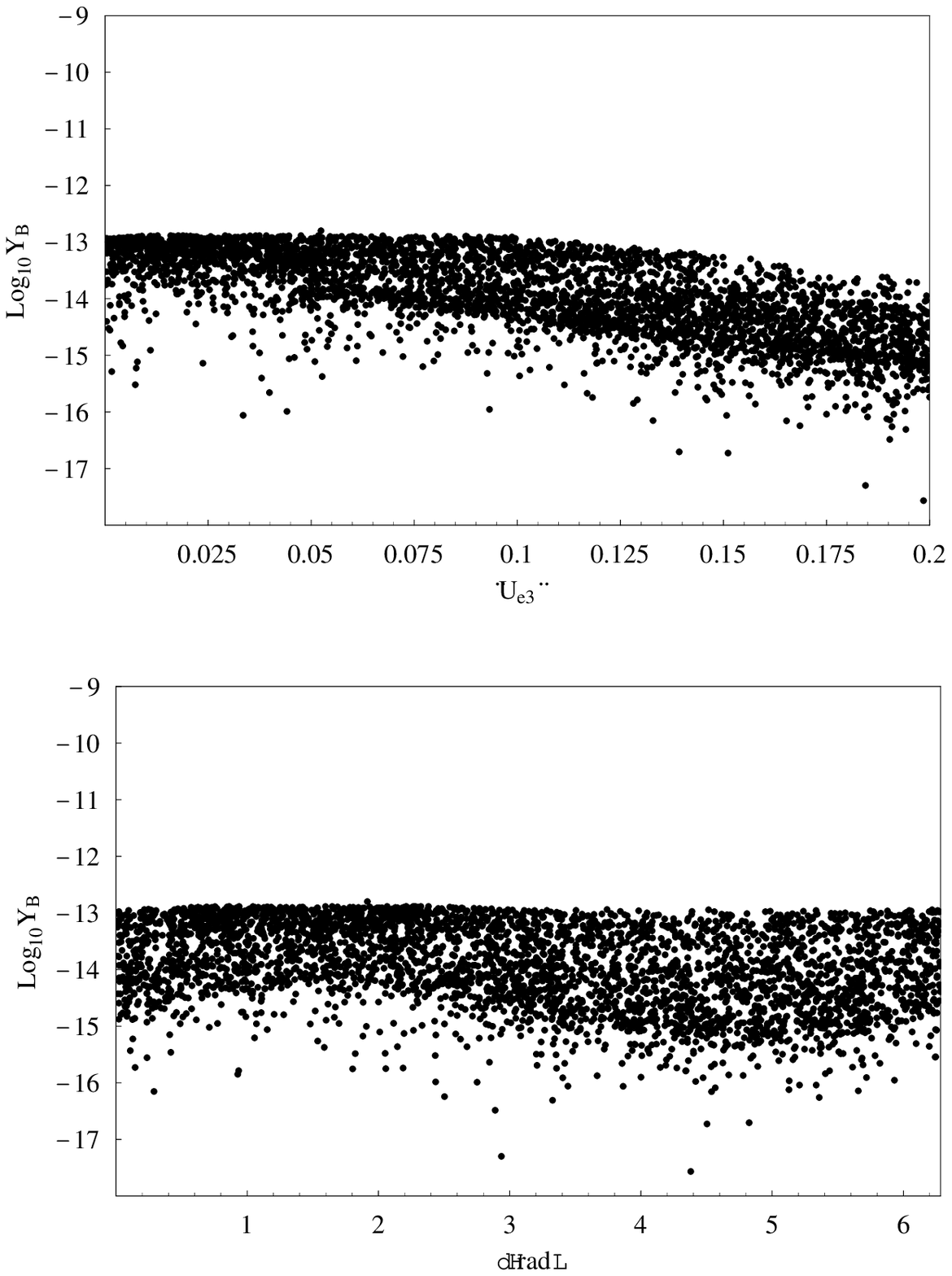,height=16cm}
\caption{The baryonic asymmetry $Y_B$ vs. $|U_{e3}|$ and $\delta$ for LMA,
quark-lepton symmetry, five texture zeros}
\end{center}
\end{figure}

\newpage

\begin{figure}[ht]
\begin{center}
\epsfig{file=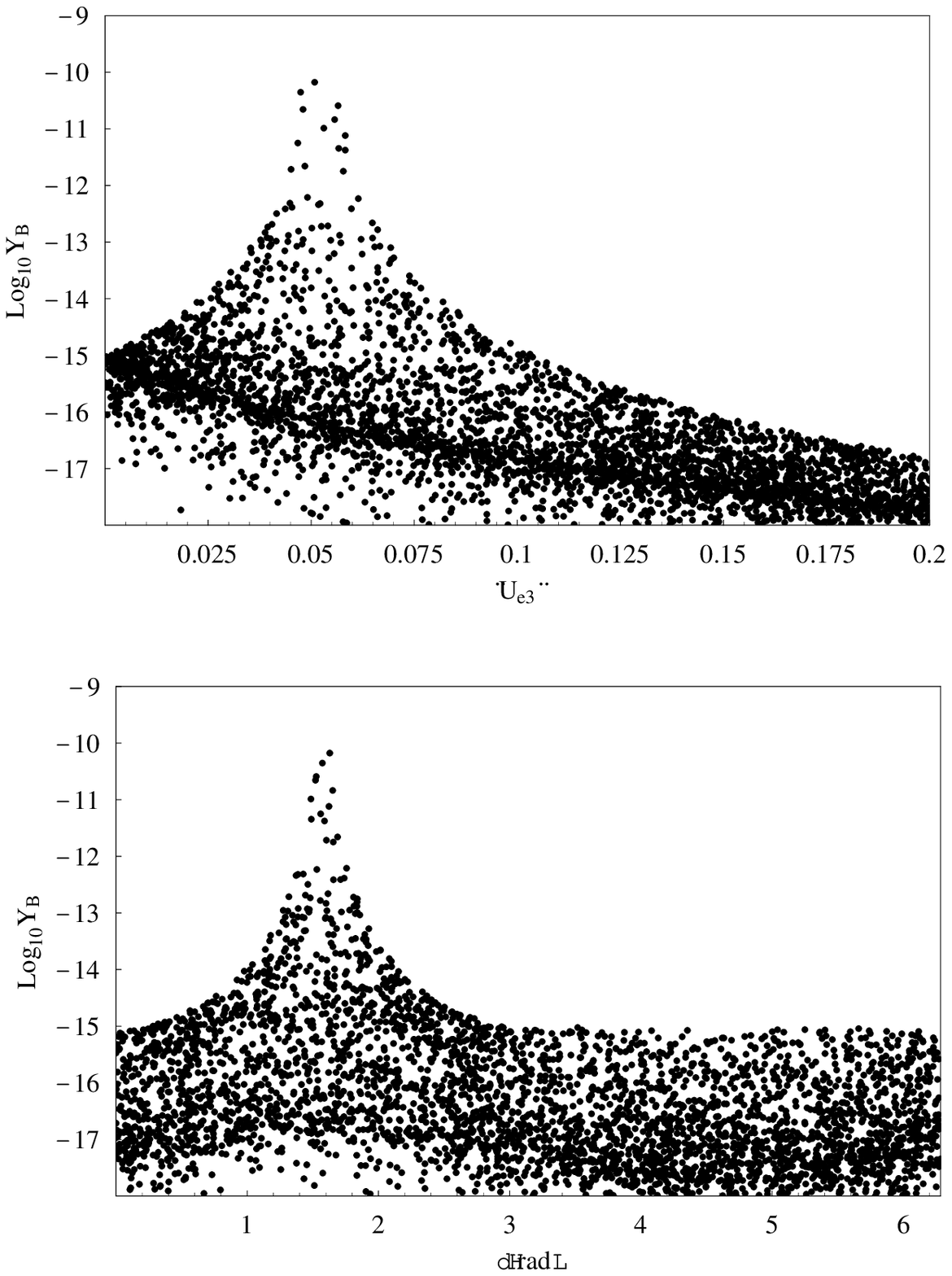,height=16cm}
\caption{The baryonic asymmetry $Y_B$ vs. $|U_{e3}|$ and $\delta$ for VO,
quark-lepton symmetry, five texture zeros}
\end{center}
\end{figure}

\newpage

\begin{figure}[ht]
\begin{center}
\epsfig{file=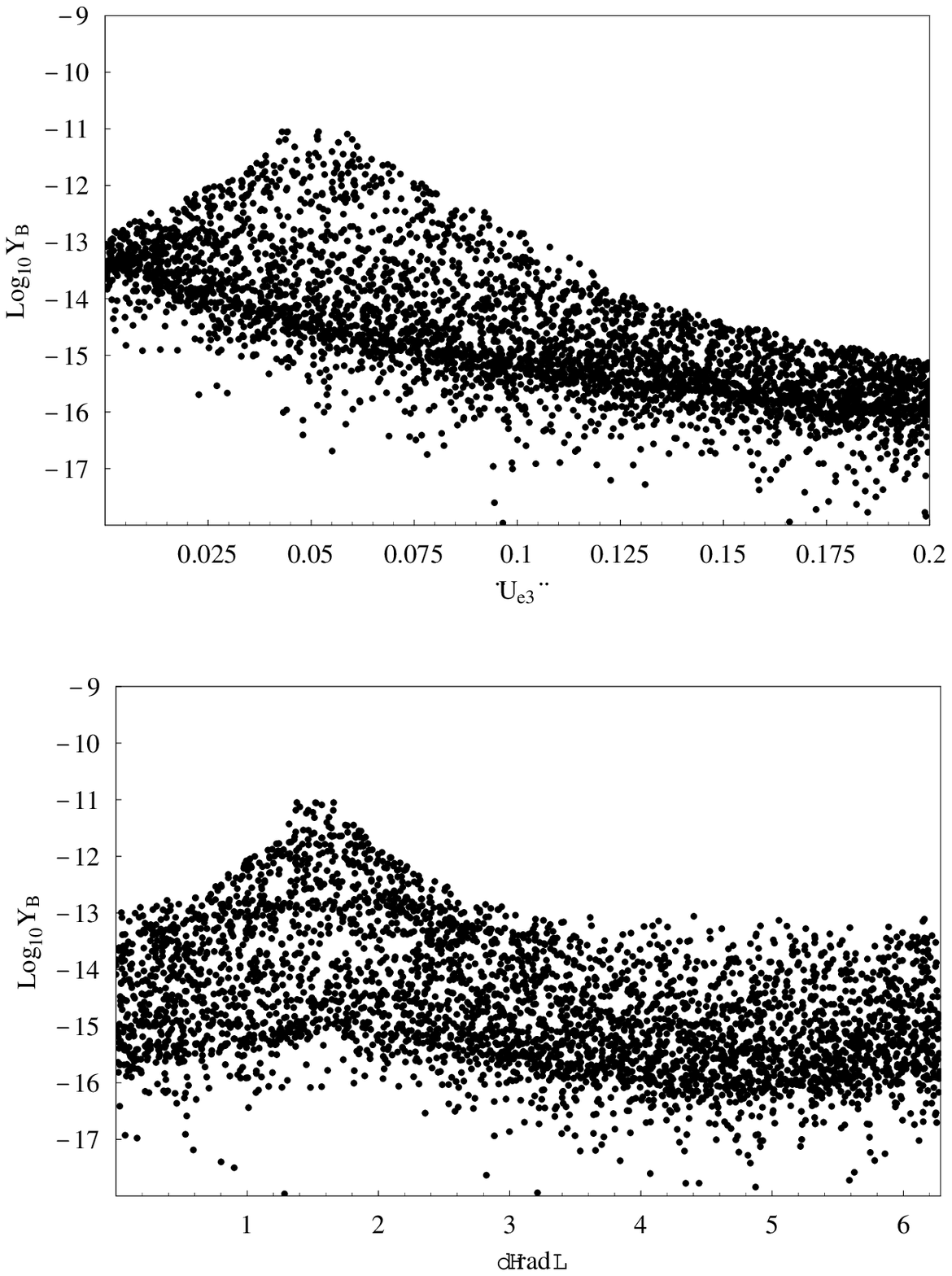,height=16cm}
\caption{The baryonic asymmetry $Y_B$ vs. $|U_{e3}|$ and $\delta$ for LOW,
quark-lepton symmetry, five texture zeros}
\end{center}
\end{figure}

\newpage

\begin{figure}[ht]
\begin{center}
\epsfig{file=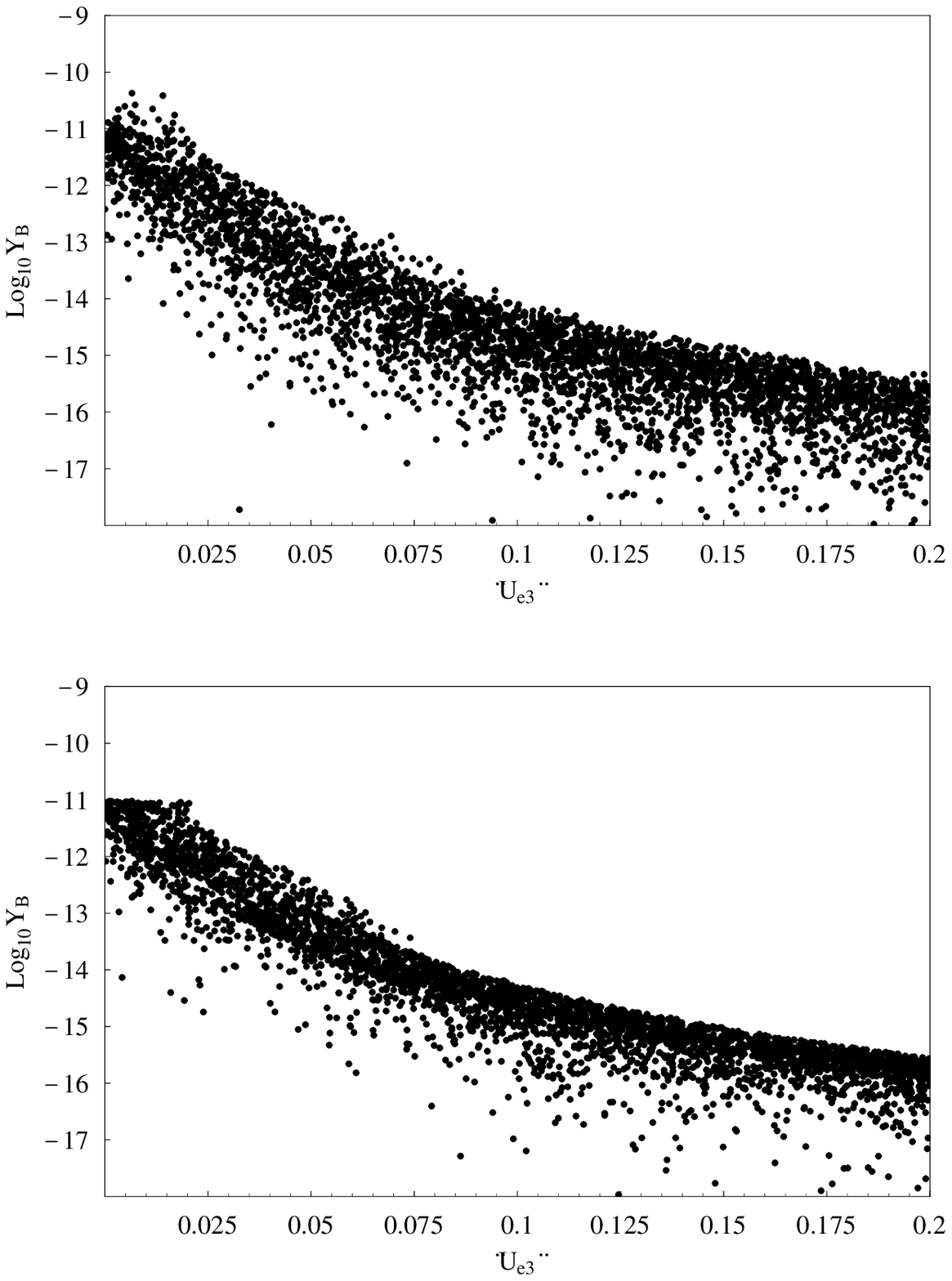,height=16cm}
\caption{The baryonic asymmetry $Y_B$ vs. $|U_{e3}|$ for SMA and LOW,
quark-lepton symmetry, four texture zeros}
\end{center}
\end{figure}

\newpage

\begin{figure}[ht]
\begin{center}
\epsfig{file=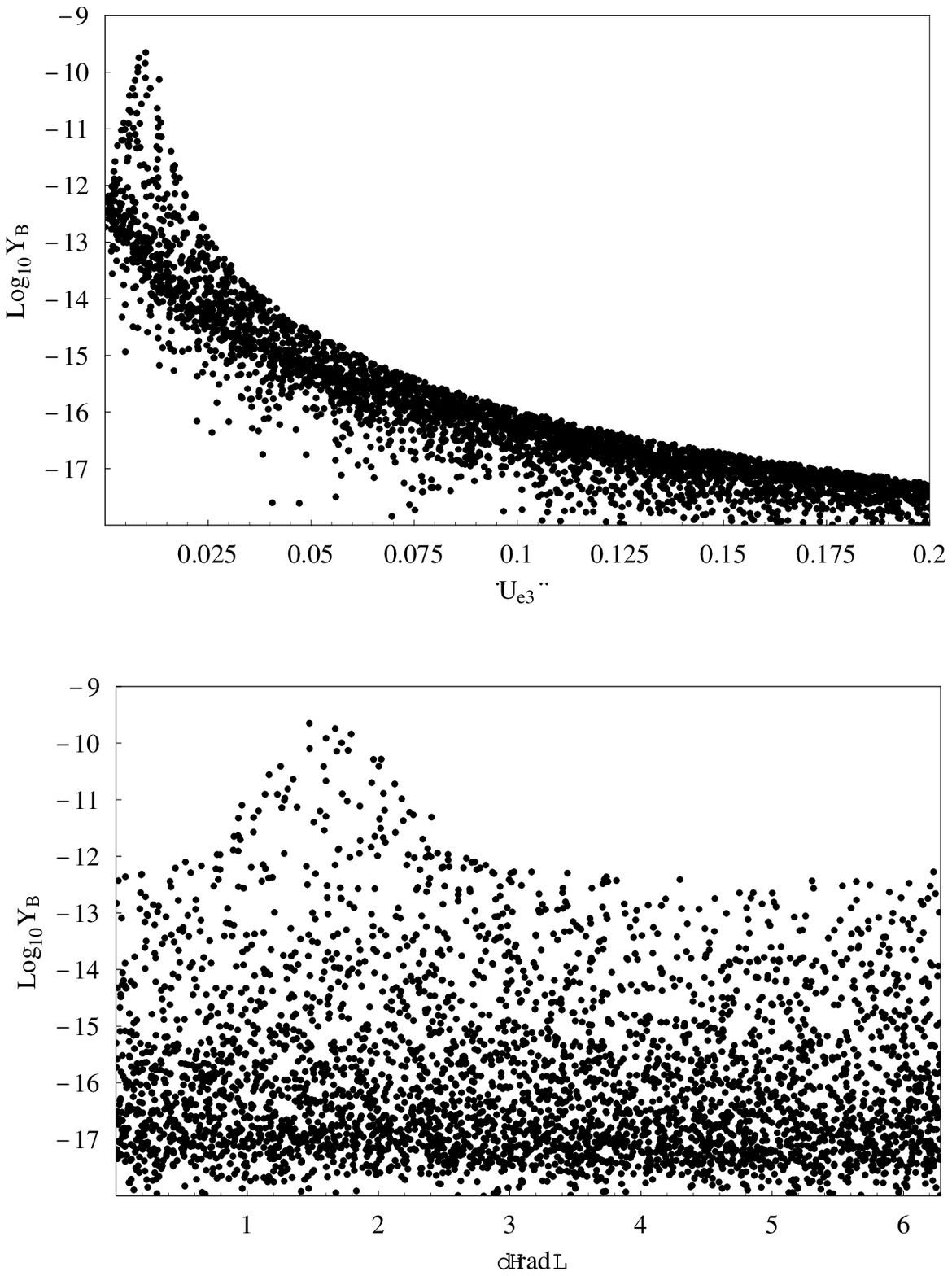,height=16cm}
\caption{The baryonic asymmetry $Y_B$ vs. $|U_{e3}|$ and $\delta$ for VO,
quark-lepton symmetry, four texture zeros}
\end{center}
\end{figure}

\newpage

\begin{figure}[ht]
\begin{center}
\epsfig{file=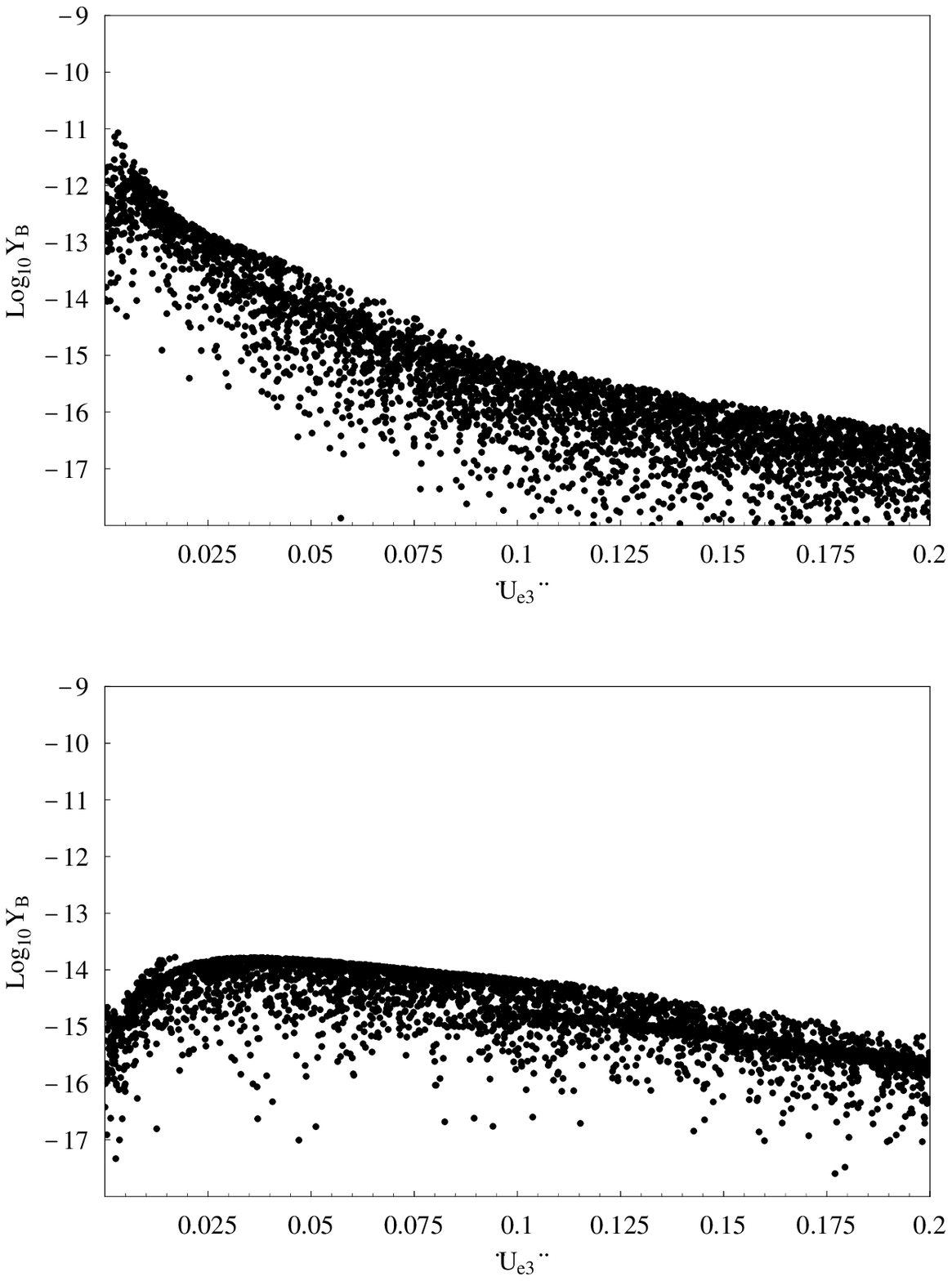,height=16cm}
\caption{The baryonic asymmetry $Y_B$ vs. $|U_{e3}|$ for SMA and LMA,
charged-neutral lepton symmetry, four texture zeros}
\end{center}
\end{figure}

\newpage

\begin{figure}[ht]
\begin{center}
\epsfig{file=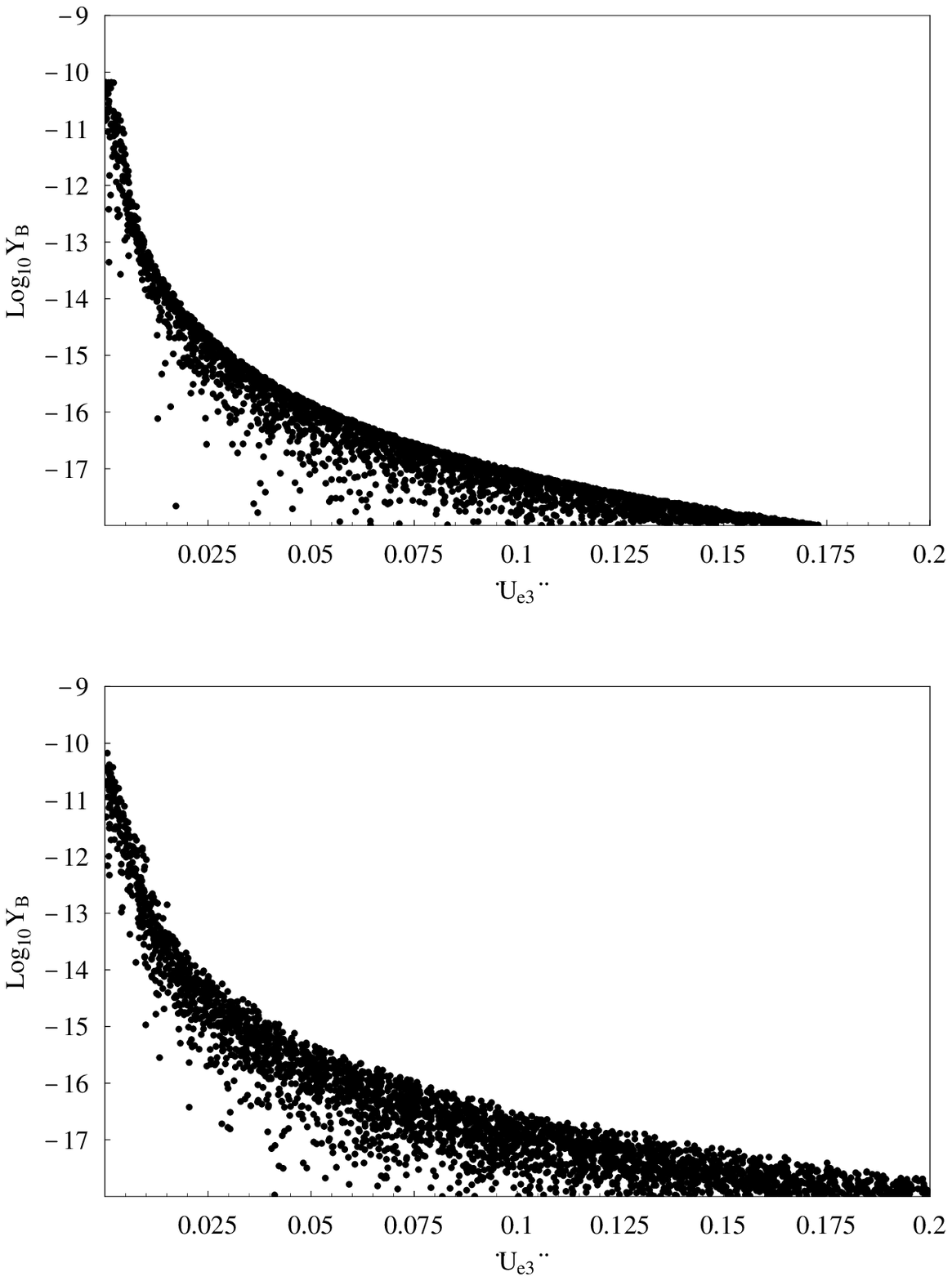,height=16cm}
\caption{The baryonic asymmetry $Y_B$ vs. $|U_{e3}|$ for VO and LOW,
charged-neutral lepton symmetry, four texture zeros}
\end{center}
\end{figure}

\end{document}